\documentclass[floatfix,reprint, superscriptaddress, amsmath,amssymb, aps]{revtex4-2}

\usepackage{pgf}
\usepackage[utf8]{inputenc}
\usepackage[T1]{fontenc}
\usepackage{graphicx}
\usepackage{dcolumn}
\usepackage{bm}
\usepackage{lipsum}
\usepackage{braket}
\usepackage{siunitx}
\usepackage{bbold}
\usepackage[colorlinks=true, linkcolor=blue, citecolor=blue, urlcolor=blue]{hyperref}
\usepackage[]{newtxtext} 
\usepackage[subscriptcorrection,nosymbolsc,smallerops,bigdelims]{newtxmath}
\DeclareMathAlphabet{\mathcal}{OMS}{cmsy}{m}{n}
\DeclareMathAlphabet{\mathbcal}{OMS}{cmsy}{b}{n}

\newcommand*{\ketbra}[2]{\ket{#1}\!\bra{#2}}
\newcommand*{\abs}[1]{\lvert #1 \rvert}
\newcommand*{\hc}{\mathrm{H.c.}}

\usepackage{lipsum}

\begin{document}

\title{Crossover from Rabi oscillations to adiabatic population switching in the Faraday optical control of quantum dot spins}

\author{Jan M. Kaspari}
\affiliation{%
 Condensed Matter Theory, TU Dortmund University, 44221 Dortmund, Germany
}%

\author{Zhe Xian Koong}
\affiliation{%
Cavendish Laboratory, University of Cambridge, JJ Thomson Ave, Cambridge CB3 0US, United Kingdom
}%
\author{Dorian A. Gangloff}
\affiliation{%
Cavendish Laboratory, University of Cambridge, JJ Thomson Ave, Cambridge CB3 0US, United Kingdom
}%
\author{Michał Gawełczyk}
%\email{michal.gawelczyk@pwr.edu.pl}
\affiliation{% 
   Institute of Theoretical Physics, Wroc{\l}aw University of Science and Technology, 50-370 Wroc{\l}aw, Poland
}%
\author{Doris E. Reiter}
\affiliation{%
  Condensed Matter Theory, TU Dortmund University, 44221 Dortmund, Germany
}%
\date{\today}

\begin{abstract}
Stimulated Raman transitions in Faraday geometry allow for simultaneous single-shot qubit readout and qubit control. It involves driving an unbalanced $\Lambda$ system via an auxiliary excited state. Due to the simultaneous driving of both transitions with unequal detuning, the resulting time-dependent Stark shift gives rise to additional resonance conditions beyond the conventional picture. We identify a distinct regime in which repeated passages through avoided crossings lead to step-like population inversion arising from Landau–Zener–St{\"u}ckelberg interference. By changing the detuning beatnote, we demonstrate a controlled continuous crossover from Rabi-like oscillations to adiabatic population switching. These findings establish the oscillating Stark shift as a mechanism for engineering and controlling spin dynamics in Faraday geometry.
\end{abstract}

\maketitle

%\section{Introduction}

{\it Introduction---}Controlling individual spins with light is a central challenge for developing scalable quantum devices and realizing efficient spin--photon interfaces \cite{Moehring2007a,Dutt2007a,Simon2007a,Gao2012,Golter2023}. Spins offer long coherence times, making them attractive candidates for quantum memories and qubits \cite{de2013quantum}. Semiconductor quantum dots provide an excellent platform for exploring spin dynamics \cite{Greilich2006,xu2009,Hogele2012a,Gangloff2019,Chekhovich2020,Shofer2025,Appel2025,Zaporski2023,Dyte2025,hogg2025fast} and simultaneously serve as bright sources of single photons \cite{Gazzano2013,Senellart2017,Holewa2022}.

\begin{figure}
    \centering
        \includegraphics[width=\columnwidth]{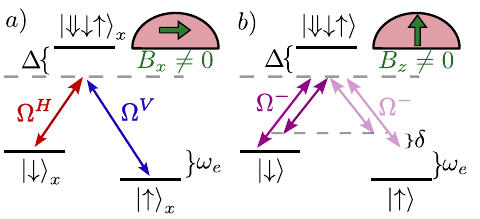}
            \caption{Energy level diagrams of the spin system in a) Voigt geometry and b) in Faraday geometry, including the spin states and the optical transitions indicated by arrows.}
            \label{fig:4LS}
\end{figure}

Optical spin control typically requires the application of magnetic fields, which we consider in a $\Lambda$ system consisting of two spin ground states and an excited state. In the first case, as shown in Fig.~\ref{fig:4LS}(a), the dipole moment of the two transitions are perpenticular to each other. This is often called Voigt geometry, where the magnetic field lies perpendicular to the optical axis and the two optical transitions form a balanced $\Lambda$ subsystem \cite{press2008complete,Bodey2019}.

Figure~\ref{fig:4LS}(b) illustrates a complementary configuration, known as Faraday geometry: here, the magnetic field is aligned along the optical axis leading to cyclic optical transitions. Faraday setups have been employed for single-shot spin readout experiments \cite{vamivakas2010observation,delteil2014observation,antoniadis2023cavity}, whose efficiency can be characterized by the cyclicity, which quantifies the relative strength of spin-conserving and spin-flipping transitions. High cyclicities thus result in strongly unbalanced $\Lambda$ systems that have made spin control difficult. As a result, Voigt configurations have served as workhorses for programmable spin rotations \cite{Bodey2019}, at the cost of single-shot readout. Recent work demonstrates that with careful compensation of differential Stark shifts both functionalities can be achieved simultaneously in Faraday-type experiments \cite{koong2025coherent}, offering new flexibility for optical spin manipulation.

% see new suggested paragraph below
% A crucial difference between Voigt and Faraday configurations lies in their transition selection rules and polarization properties, as indicated in Fig.~\ref{fig:4LS} by the double-sided arrows. In Voigt geometry, laser polarization can be chosen to selectively address either spin-conserving via horizontally polarized light $\Omega^H$ or spin-flipping transitions via horizontally polarized light $\Omega^V$. In contrast, in Faraday geometry both transitions are excited simultaneously due to their selection rules with circularly polarized light $\Omega^-$, only the strength differs. This simultaneous driving with unequal detunings from the two transitions induces a time-dependent differential AC Stark shift of the spin states, dynamically modifying their energy splitting. Consequently, while well-defined two-photon resonances exist in Voigt systems, in Faraday geometry the resonance structure is no longer determined solely by static laser detunings, but is governed by the dynamical modulation of the spin splitting.

% Suggested replacement for the above paragraph
A crucial difference between Voigt and Faraday configurations lies in the selection rules and polarization properties of their respective $\Lambda$ systems, as indicated in Fig.~\ref{fig:4LS} by the double-sided arrows. Typical spin control employs a pair of identically polarised Raman lasers with an energy difference matching the ground state splitting~\cite{Bodey2019}. In Voigt geometry, the balanced $\Lambda$ system allows one to select circularly polarised Raman lasers resulting in equal coupling to the spin-conserving leg, via the horizontally polarized component $\Omega^H$, and spin-flipping leg, via the vertically polarized component $\Omega^V$. This configuration avoids differential Stark shifts on the ground state during spin control~\cite{Jackson2021}. By contrast, in Faraday geometry it is possible for both spin-conserving and spin-flipping transitions to have the same circularly-polarised transition dipole, owing to a particular kind of hole mixing, only the oscillator strength differs~\cite{koong2025coherent}. In this situation, the resonance structure acquires a richer structure as it is no longer determined solely by static laser detunings, but is governed by the dynamical modulation of the spin splitting.  

In this paper, we revisit theoretically the resonances of co-polarised imbalanced $\Lambda$ systems, specifically those of quantum dots in Faraday geometry. We identify a multitude of resonances induced by the oscillating differential Stark shifts arising from the beatnote of the two Raman lasers simultaneously, but unequally, driving the two optical transitions. Among conventional resonances such as the regular two-photon resonance, we identify regimes involving the exchange of multiples of the beatnote frequency. In contrast to the conventional resonances that produce sinusoidal Rabi oscillations, we find spin control in  unprecedented regimes that is achieved via Landau-Zener-St{\"u}ckelberg interference, when the differential AC Stark shift repetitively pulls the spin states through an avoided crossing. These new resonances can be tuned from the diabatic limit leading to Rabi-like rotation through intermediate step-like population transfer dynamics, and all the way to the adiabatic limit with rapid periodic switching of the spin state. Such behavior opens up promising routes toward novel schemes for coherent optical spin control.

{\it System and model---}%
We consider a singly charged quantum dot with a resident electron, subject to a magnetic field in Faraday geometry, and coupled to negative trion states via Raman laser fields as sketched in Fig.~\ref{fig:4LS}(b). Given the selection rules in the system, circularly polarized laser beams select a three-level system in a $\Lambda$ configuration, as shown in Fig.~\ref{fig:4LS}(b) for the case of left-circular polarization $\sigma_-$ considered throughout the paper. Note that in principle, the full system consists of four levels, however, due to our choice of polarization, we can restrict ourselves to three levels only. The two lower levels are the electron spin-down $\ket{\downarrow}$ and spin-up $\ket{\uparrow}$ states, separated by the Zeeman energy $\hbar\omega_{\mathrm{e}}$. Both states are optically coupled to the trion state $\ket{T}=\ket{\Downarrow\uparrow\downarrow}$ at energy $\hbar\omega_{\mathrm{T}}$ relative to the spin-down state. The transition $\ket{\downarrow}\leftrightarrow\ket{T}$ is angular-momentum-allowed and nominally bright, while $\ket{\uparrow}\leftrightarrow\ket{T}$ is nominally dark and can become brightened in a QD due to the mixing of the heavy and light hole subbands \cite{koong2025coherent}. Thus, both transitions are driven simultaneously by the optical field, while the corresponding transition Rabi rates differ. Their ratio is given by the imbalance $\eta$. Note that the imbalance $\eta$ is connected to the cyclicity $C$ via $\eta=\sqrt{C}$. A hypothetical balanced system would have $\eta=1$, whereas realistic Faraday spin systems typically exhibit $\eta\gg1$.
The system is driven by two $\sigma_-$-polarized continuous-wave laser fields, 
\begin{equation}
    \bm{E}(t) =\left[ E_1^{-} e^{-i\omega_1 t} + E_2^{-} e^{-i\omega_2 t} + \mathrm{c.c.} \right] \bm{e}_{{\sigma}_{-}}\,.
\end{equation}
The first laser is detuned by $\Delta$ from the spin-conserving optical transition, $\omega_1=\omega_{\mathrm{T}}-\Delta$, while the second has a frequency mismatch or beatnote $\delta$ relative to the first, $\omega_2=\omega_1+\delta$, with $|\delta|\ll|\Delta|$. For simplicity, we take both laser fields to have the same amplitude, $E_1^-=E_2^- =E$.

Within the dipole and rotating wave approximations, the full Hamiltonian of the system in the rotating frame set to the laser frequency of pulse one with $\omega_{\mathrm{L}}=\omega_{\mathrm{T}}-\Delta$, reads
\begin{align}
    {H}_{\mathrm{RWA}}(t) ={}& -\hbar\omega_{\mathrm{e}} \ketbra{\uparrow}{\uparrow} + \hbar\Delta \ketbra{T}{T} \nonumber \\
    &{}- \frac{\hbar}{2} \left[\widetilde{\Omega}(t)\left( \ketbra{T}{\downarrow} + {\frac{1}{\eta}}\ketbra{T}{\uparrow} \right) + \hc \right],
\label{eq:three}
\end{align}
where the driving term carries the beatnote $\delta$ of the two lasers,
\begin{equation}
    \widetilde{\Omega}(t) = \Omega + \Omega e^{-i\delta t} = 2 \Omega \cos\left( \frac{\delta}{2}t \right) e^{-\frac{1}{2} i\delta \, t} \label{eq:lightfield}\,.
\end{equation}
Here, $\Omega$ is proportional to $dE$, where we have assumed the dipole moment of the spin-conserving transition to be $\bm{d} = d_0\bm{e}_{{\sigma}_{-}}$.

{\it Analytical treatment---}% spin model---}%
To obtain analytical resonance conditions for efficient spin control, we start by deriving an effective model. Working in the regime of $\Delta\gg\Omega$, we can exploit the separation of energy- and thus time-scales in the trion evolution, and perform an adiabatic elimination to reduce the problem to an effective two-level spin model \cite{Brion2007}
with an effective Hamiltonian written in terms of $\ket{\uparrow/\downarrow}$ basis Pauli matrices $\sigma_i$
\begin{equation}\label{eq:effTLS}
    H_{\mathrm{eff}}(t) = \frac12 \hbar\omega_{\mathrm{diff}}(t) \, \sigma_z - \frac12\hbar\Omega_{\mathrm{spin}}(t) \, \sigma_x \,,
\end{equation}
where we have performed a global time-dependent energy shift $\eta\hbar\Omega_{\mathrm{spin}}(t)/2$. We introduced the splitting $\omega_{\mathrm{diff}}(t) = \omega_{\mathrm{S}}(t) - \omega_{\mathrm{e}}$ containing the differential AC Stark shift
\begin{align}\label{eq:effStark}
    \omega_{\mathrm{S}}(t) = \frac{\abs{\widetilde{\Omega}(t)}^2}{4\Delta}\left(1-\frac{1}{\eta^{2}}\right)
                        = \frac{\Omega^2
                                }{\Delta}
                                \cos^2\left( \frac{\delta}{2}t\right)
                                \left( 1-\frac{1}{\eta^{2}} \right),
\end{align}
as well as the effective driving %Rabi frequency 
\begin{equation}
    \Omega_{\text{spin}}(t) = {\frac{ \abs{\widetilde{\Omega}(t)}^2}{2\eta\Delta}} 
    = {\frac{2 \Omega^2}{\eta\Delta}}\cos^2 \left(\frac{\delta}{2}t\right) \,. 
    \label{eq:effRabi} 
\end{equation}
Because of the frequency mismatch $\delta$ between the two laser fields [cf. Eq.~\eqref{eq:lightfield}], the differential Stark shift, and thus the effective spin splitting are modulated at beatnote $\delta$. It is remarkable to note that this only happens for $\eta > 1$, while in a balanced system $\omega_{\mathrm{S}}=0$. The time dependence of both driving $\Omega_{\text{spin}}$ and splitting $\omega_{\text{diff}}$ implies that resonance conditions are no longer determined by static detuning, but by dynamical modulation of the spin splitting.

%{\it Analytical resonance conditions---}%
To determine the resonance conditions leading to spin rotations,
we look for secular terms that could drive such evolution.
To this end, we shift all time dependence to the driving term through the unitary transformation $\widetilde{H} = UHU^\dag+i\hbar\dot{U}U^\dag$ with
\begin{align}
    U(t) ={}& \exp\left({i\!\int_0^t\!\!\mathrm{d}\tau \, \frac{\omega_{\mathrm{diff}}(t)}{2} \, \sigma_z}\right) \nonumber \\
    ={}& \exp\left(\frac{i}{2}\left[ \chi\delta t +\chi\sin(\delta t)-\omega_{\mathrm{e}}t\right]\,\sigma_z\right),
\end{align}
where we have defined
\begin{equation}
    \chi=\frac{\Omega^2}{2\Delta\delta}\left(1-\frac{1}{\eta^2}\right) \,.
\end{equation}
With this transformation, we eliminate the splitting, such that the transformed effective Hamiltonian reads
\begin{equation}\label{eq:transformed_effective_hamiltonian}
    \widetilde{H}_{\text{eff}} = -\frac{\hbar}{2} \Omega_{\text{spin}}(t)\,e^{i\chi \delta t -i\omega_{\mathrm{e}} t + i \chi \sin(\delta t)} \ketbra{\downarrow}{\uparrow} + \hc
\end{equation}

Next, we carry out a Jacobi-Anger expansion $e^{i\chi\sin(\delta t)} = \sum_{n=-\infty}^{\infty} J_n(\chi)\, e^{in\delta t}$ to express Eq.~\eqref{eq:transformed_effective_hamiltonian} in the harmonics of the beatnote $\delta$, where $J_n$ is the $n$-th Bessel function of the first kind.
Combining this with Eq.~\eqref{eq:effRabi}, we obtain
\begin{align}\label{eq:final-eff-ham}
    \widetilde{H}_{\text{eff}} = -\frac{\hbar}{2} \sum_{n=-\infty}^{\infty} \widetilde{\Omega}_{\text{spin}}^{(n)} \, e^{i\left[\delta\left(\chi+n\right)-\omega_{\mathrm{e}}\right]t} \, \ketbra{\downarrow}{\uparrow} + \hc,
\end{align}
where, assuming the frequencies are sufficiently spaced out ($\chi\ll n$), for each harmonic order $n$, a secular term dominating the evolution can be found when
\begin{equation}
    \delta (\chi+n)  = \frac{\Omega^2}{2\Delta}\left(1-\frac{1}{\eta^2}\right) + \delta \, n= \omega_{\mathrm{e}} \,.
    \label{eq:resonance}
\end{equation}
The corresponding effective Rabi frequency is
\begin{align}\label{eq:transformed_effective_rabi}
     \widetilde{\Omega}_{\text{spin}}^{(n)} ={}& \frac{\Omega^2}{2\eta\Delta} \left[ 2{J}_{n-1}(\chi) + {J}_n(\chi) + {J}_{n+1}(\chi)\right].
\end{align}

Equation~\eqref{eq:final-eff-ham} can be interpreted as a Floquet expansion, where the resonance condition in Eq.~\eqref{eq:resonance} corresponds to quasi-energy degeneracies between states differing by $n$ drive quanta \cite{Bukov2015}. Note that if multiple harmonics satisfy the resonance condition within an effective coupling bandwidth, the independent-drive picture will break down, leading to interference effects beyond simple Rabi dynamics, as discussed below.

%\section{Results}
{\it Results---}%
Equipped with Eq.~\eqref{eq:resonance}, which analytically predicts the conditions for effective resonant Rabi drives in the spin system, we proceed to study different cases via simulations. For this, we numerically solve the Liouville-von Neumann equation, $\dot{\rho}(t)=-i\left[H(t),\rho(t)\right]/\hbar$, for the density matrix $\rho$ with the Hamiltonian from Eq.~\eqref{eq:three}. The system parameters are taken from a recent experiment on a single semiconductor quantum dot \cite{koong2025coherent}: $\eta = 20.28$, $\hbar \Delta = 2.481~\mathrm{meV}$ ($600$~GHz), $\hbar\omega_{\mathrm{e}} = 11$~\textmu{}eV ($2.6$~GHz), while $\Omega$ and $\delta$ are tuned in the different cases. All figures that follow show the numerical results from our full three-level model.

\begin{figure}
    \centering
        \includegraphics[width=1\columnwidth]{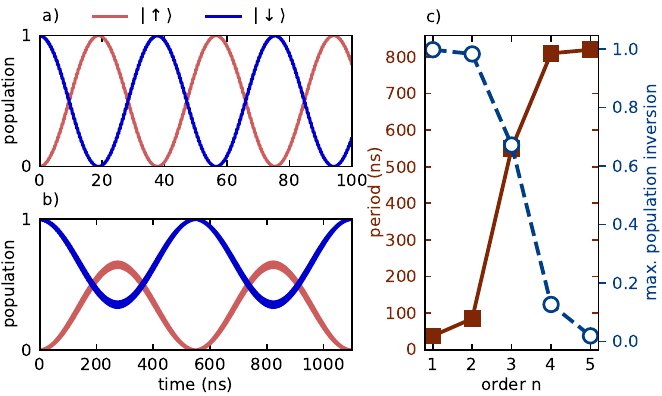}
            \caption{Population dynamics of the spin states for $\Omega=23.87$~GHz for a) $n=1$ with $\hbar\delta=8.79$~\textmu{}eV ($2.13$~GHz) and b) $n=3$ with $\hbar\delta=2.93$~\textmu{}eV ($0.71$~GHz). c) Period of the oscillations (red) and the maximal achievable population inversion (blue) for orders $n=1,\dots,5$. The lines are just a guide to the eye. 
            }
            \label{fig:unbalanced}
\end{figure}

In the case $n\neq 0$, the resonance condition Eq.~\eqref{eq:resonance} imposes a constraint on $\delta$. This is expected, as the process underlying the evolution involves the exchanges of multiples of that beatnote to drive the spin system.

Figure~\ref{fig:unbalanced} shows the population dynamics of the spin states $\ket{\uparrow}$ and $\ket{\downarrow}$. At the fundamental resonance [$n=1$, Fig.~\ref{fig:unbalanced}(a)], the spin undergoes coherent rotations with periodic complete population inversion. On top of that, weak high-frequency oscillations due to nonsecular terms are visible. At a higher-order resonance [$n=3$, Fig.~\ref{fig:unbalanced}(b)],
we also observe coherent oscillations, but on a longer timescale and with more pronounced higher frequency components. Complete inversion is not reached because the secular approximation after the Jacobi-Anger expansion does not hold as strictly anymore.
Note that this result is obtained exactly at the analytical resonance condition, and that higher inversion may be achievable by optimizing parameters to compensate for deviations from the effective model.
As expected, increasing $n$ leads to an increase in the oscillation period as seen in Fig.~\ref{fig:unbalanced}(c), reflecting the reduced effective coupling strength. In addition,
for larger $n$, the reduced spacing between harmonics perturbs the evolution with nonsecular drives that are not fast enough to average out, causing the numerical result to deviate from analytical predictions with amplitude $1$. Changing the laser power primarily affects the effective Rabi frequency $\Omega_{\mathrm{spin}}$, resulting in a corresponding change in the oscillation period.

\begin{figure}
    \centering
        \includegraphics[width=\columnwidth]{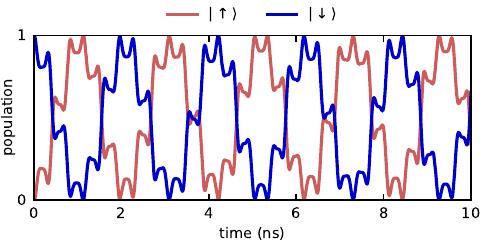}
            \caption{Population dynamics of spin states for $n=1$ in the balanced case $\eta=1$ with $\Omega = 23.87$~GHz and $\hbar\delta=10.75$~\textmu{}eV ($2.6$~GHz).}
            \label{fig:balanced}
\end{figure}

We briefly want to compare our findings to the balanced case $C=1$, in which the resonance condition reduces to
\begin{equation}
    \delta = \frac{\omega_{\mathrm{e}}}{n}     \,.
\end{equation}
In Voigt geometry, $n=-1$ is known as the two-photon Raman process \cite{yatsiv1968enhanced,braunlich1970detection} and $|n|>1$ are higher harmonics. While in the Voigt case with circularly polarised lasers, a clear sin-modulation is expected, the population dynamics in Faraday configuration, as shown in Fig.~\ref{fig:balanced}, shows an irregular oscillation. Only the mean value follows a sine function. The additional oscillation can be traced to the fact that the two laser frequencies always act on both transitions and the resulting effective Rabi frequency $\Omega_{\text{spin}} \sim \cos^2\left( \tfrac{\delta}{2}t\right)$ therefore oscillates with the frequency mismatch $\delta$.

\begin{figure*}[ht]
    \centering
        \includegraphics[width=1.0\textwidth]{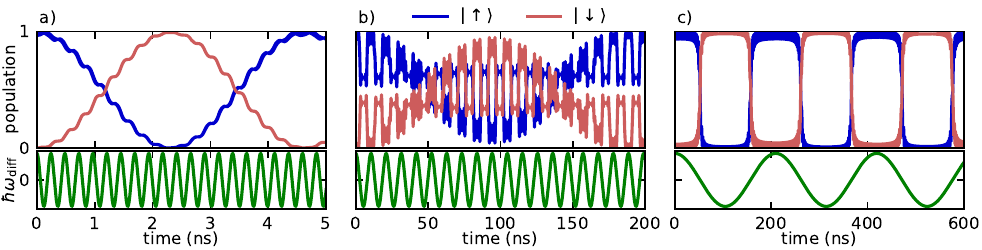}
            \caption{Population dynamics of spin states (top panels) and instantaneous diabatic energy difference $\hbar\omega_{\text{diff}}$ (bottom) for $n=0$ at a) diabatic limit: $\hbar\delta=17.11$~\textmu{}eV ($4.14$~GHz), b) intermediate step-like dynamics: $\hbar\delta=0.39$~\textmu{}eV ($0.10$~GHz), and c) adiabatic limit: $\hbar\delta=19.75$~neV ($4.77$~MHz). 
            }
            \label{fig:n=0}
\end{figure*}

We now turn to the exceptional case of $n=0$, which defines a distinct resonance condition given by
\begin{align}\label{eq:condition-n-0}
    \frac{\Omega^2}{2\Delta}\left(1-\frac{1}{\eta^2}\right) = \omega_{\mathrm{e}} \,. %\,\,\,
\end{align}
It fixes the laser amplitude $\Omega$, while leaving the pulse detuning $\delta$ free, as the drive is now secular for any $\delta$. This is understandable, as $n=0$ suggests that the mechanism of this resonance is completely different from those considered above and does not involve exchanges of the beatnote quanta.

Figures~\ref{fig:n=0}(a--c) show the corresponding dynamics of spin state populations for three different laser beatnotes. In Fig.~\ref{fig:n=0}(a), we observe nearly sinusoidal population oscillations with a small superimposed higher-frequency modulation. As $\delta$ is reduced, qualitatively different dynamics emerge in Fig.~\ref{fig:n=0}(b), where a staircase-like sine behavior is found, and 
at even lower
$\delta$, in Fig.~\ref{fig:n=0}(c), where the populations exhibit complete switching at discrete times, while remaining approximately constant otherwise.
This set of behaviors resembles, on one end, a regular Rabi rotation; on the other end, the Landau-Zener-St{\"u}ckelberg (LZS) interference through the intermediate interference case, transitioning to the adiabatic (strong-coupling) limit \cite{Shevchenko2010}. 
Inspecting again Eqs.~\eqref{eq:effTLS}-\eqref{eq:effRabi}, the instantaneous diabatic energy difference  for $n=0$ reduces to
\begin{equation}
    \hbar\omega_{\mathrm{diff}}(t) = \hbar\omega_{\mathrm{e}} \cos(\delta t) \,,
\end{equation}
while the off-diagonal driving term becomes
\begin{equation}
    \Omega_{\mathrm{spin}}(t) =
    \frac{2\eta\omega_{\mathrm{e}}}{\eta^2-1} 
    \left[1+\cos(\delta t)\right] 
    \approx2\frac{\omega_{\mathrm{e}}}{\eta}  \left[1+\cos(\delta t)\right],
\end{equation}
i.e., we indeed repeatedly, at times $t_m = (2m+1)\pi/(2\delta)$ (separated by $\tau=\pi/\delta$), drive the system through the avoided crossing set by the effective constant spin coupling $2\omega_{\mathrm{e}}/\eta$. The coupling also has an additional component that oscillates in phase with the splitting.

At each $t_m$, to leading order, the splitting can be linearized as $\omega_{\mathrm{diff}}(t)\approx \pm\omega_{\mathrm{e}}\delta (t-t_m)$, and the coupling becomes $\Omega_{\text{spin}}\approx2\omega_{\mathrm{e}}/\eta$. The system undergoes Landau-Zener crossings with sweep velocity $v=\abs{\dot{\omega}_{\mathrm{diff}}}_{t_m}=\omega_{\mathrm{e}}\delta$ and anticrossing width $4\omega_{\mathrm{e}}/\eta $. This gives the probability of remaining in the current diabatic state ($\ket{\uparrow{/}\downarrow}$ basis) of $P=\exp(-2\pi\gamma)$ with the Landau-Zener adiabaticity parameter
\begin{equation}\label{eq:adiabaticity-gamma}
    \gamma = \frac{\Omega_{\mathrm{spin}}^2(t_m)}{4v} \approx\frac{\omega_{\mathrm{e}}}{\delta\eta^2},
\end{equation}
where we approximated for $\eta\gg1$. Each crossing (partly) flips the spin and adds a relative phase, while between crossings the evolution is adiabatic despite the additional oscillatory coupling, and accumulates additional relative phase.

Approaching the diabatic limit of $\gamma\to 0$ ($\delta\gg\omega_{\mathrm{e}}/\eta^2$) can lead to two types of dynamics, each producing Rabi-like behavior at different rates. If the beatnote is fast compared with spin dynamics, while within the validity range of the effective model, we get $\overline{\cos(\delta t)}\approx 0$ and the spin effectively experiences the time-averaged
Hamiltonian $\overline{H}_{\mathrm{eff}}\approx-(\hbar\omega_{\mathrm{e}}/\eta)\sigma_x$. For smaller but still diabatic beatnotes, one recovers the standard LZS fast-passage limit as shown in Fig.~\ref{fig:n=0}(a). In this case, we deal with tiny occupation transfers at each passage, which accumulate to generate a quasi-continuous Rabi-like behavior. 

At the other extreme of $\gamma\to\infty$ ($\delta\ll\omega_{\mathrm{e}}/\eta^2$), realized in Fig.~\ref{fig:n=0}(c), we deal with the adiabatic limit in which $P\to0$. Now, at each crossing of $\omega_{\mathrm{diff}}= 0$, a full spin flip occurs, a phenomenon also known as the adiabatic rapid passage \cite{malinovsky2001general} used, e.g., for population inversion of charge states in quantum emitters~\cite{simon2011robust,wu2011population,kappe2025chirped}.
When repeated periodically, it leads to spin evolution in which populations no longer transfer continuously but consist of fast flips at each $t_m$, producing a square-like behavior. 
Note that $H_{\mathrm{eff}}(t)$ is dynamically self-similar under the simultaneous scaling of $\omega_{\mathrm{e}}$ and $\delta$, which corresponds to the scaling of time. Thus, the achievable Zeeman splitting sets an upper bound on the spin-flipping frequency, with an improvement of more than an order of magnitude possible for a mixed magnetic field \cite{SchimpfPRXQ2025}.

It may appear surprising that LZS dynamics emerge from Eqs.~\eqref{eq:final-eff-ham} and \eqref{eq:resonance}, which predict a set of independent, resonant Rabi-like drives associated with distinct harmonic orders. For $n \neq 0$, the resonance condition selects an essentially unique $\delta$, corresponding to a well-isolated harmonic. In contrast, the $n = 0$ condition is qualitatively different: for any fixed $\delta$, it can lie close to many higher-order $n\neq0$ resonances, whose contributions depend both on their detuning from resonance, $|\delta(\chi + n) - \omega_e|$, and their amplitude, as given by Eq.~\eqref{eq:transformed_effective_rabi}. Thus, another view of the crossover from Rabi-like to switching behavior in our LZS spin control is that it arises from a buildup of multiple harmonics at low $\delta$, which collectively shape the dynamics and lead to increasingly square-like population behavior.

%\section{Conclusion}

{\it Conclusions---}%
We have investigated optical spin control in a Faraday $\Lambda$ system and identified distinct sets of resonance conditions arising from the interplay between coherent optical driving and the induced time-dependent differential AC Stark shift. Using an effective two-level description, we showed that the oscillating Stark shift produces a periodically modulated energy splitting, fundamentally altering the resonance structure compared to conventional Raman schemes governed by static detuning.

For unbalanced systems ($\eta \neq 1$), this dynamical modulation gives rise to a family of resonance conditions associated with different harmonic orders $n$, enabling coherent population transfer despite a strong imbalance between the optical transitions. In addition, we identified a distinct regime corresponding to $n=0$, where the resonance condition fixes the laser amplitude rather than the detuning. In this regime, repeated passages through avoided crossings induced by the oscillating Stark shift give rise to spin dynamics governed by Landau–Zener–St{\"u}ckelberg interference and characterized by step-like population inversions.

Such behavior occurs under two conditions in a $\Lambda$-system: the driving laser pulses must be at least partially co-polarized (i.e., $\delta \neq 0$), and there must be an imbalance $\eta > 1$. In addition, the AC Stark effect needs to be pronounced enough. While our discussion has focused on two distinct cases found in semiconductor quantum dots, similar conditions may also arise under asymmetric driving \cite{samaras2026} or in other systems, such as vacancy centers in diamond \cite{Pingault2017, Debroux2021, Hepp2014}.

Our results establish the oscillating differential AC Stark shift as a mechanism for dynamically engineering spin resonance conditions and controlling spin dynamics in Faraday geometry, thereby extending the capabilities of optical spin manipulation in solid-state quantum systems.

%\section*{Acknowledgement}
{\it Acknowledgements---}%
J.M.K. and D.E.R. gratefully acknowledge funding through the QuantERA Project MEEDGARD from the German Federal Ministry of Research, Technology and Space (BMFTR; grant number 16KIS2058).
M.G.\ acknowledges the financing of the MEEDGARD project funded within the QuantERA II Program that has received funding from the European Union's Horizon 2020 research and innovation program under Grant Agreement No.\ 101017733 and the National Centre for Research and Development, Poland -- project No.\ QUANTERAII/2/56/MEEDGARD/2024. D.A.G and Z.X.K acknowledge support from an EPSRC New Investigator Award EP/W035839/2 and the QuantERA project MEEDGARD through EPSRC EP/Z000556/1. D.A.G. acknowledges support from a Royal Society University Research Fellowship.
% \appendix
% %
% \section{Equations for the pulses}
\bibliography{PaperBib.bib}

@PREAMBLE{
 "\providecommand{\noopsort}[1]{}" 
 # "\providecommand{\singleletter}[1]{#1}%" 
}

@article{press2008complete,
  title={Complete quantum control of a single quantum dot spin using ultrafast optical pulses},
  author={Press, David and Ladd, Thaddeus D and Zhang, Bingyang and Yamamoto, Yoshihisa},
  journal={Nature},
  volume={456},
  number={7219},
  pages={218--221},
  year={2008},
  url={https://www.nature.com/articles/nature07530}
}

@incollection{de2013quantum,
  title={Quantum Memories: Quantum Dot Spin Qubits},
  author={De Greve, Kristiaan},
  booktitle={Towards Solid-State Quantum Repeaters: Ultrafast, Coherent Optical Control and Spin-Photon Entanglement in Charged InAs Quantum Dots},
  pages={25--38},
  year={2013},
  publisher={Springer}
}

@article{vamivakas2010observation,
  title={Observation of spin-dependent quantum jumps via quantum dot resonance fluorescence},
  author={Vamivakas, A Nick and Lu, C-Y and Matthiesen, C and Zhao, Ya and F{\"a}lt, S and Badolato, A and Atat{\"u}re, M},
  journal={Nature},
  volume={467},
  number={7313},
  pages={297--300},
  year={2010},
  url={https://www.nature.com/articles/nature09359}
}

@article{delteil2014observation,
  title={Observation of quantum jumps of a single quantum dot spin using submicrosecond single-shot optical readout},
  author={Delteil, Aymeric and Gao, Wei-bo and Fallahi, Parisa and Miguel-Sanchez, Javier and Imamo{\u{g}}lu, Atac},
  journal={Phys. Rev. Lett.},
  volume={112},
  number={11},
  pages={116802},
  year={2014},
  url={https://journals.aps.org/prl/abstract/10.1103/PhysRevLett.112.116802}
}

@article{antoniadis2023cavity,
  title={Cavity-enhanced single-shot readout of a quantum dot spin within 3 nanoseconds},
  author={Antoniadis, Nadia O and Hogg, Mark R and Stehl, Willy F and Javadi, Alisa and Tomm, Natasha and Schott, R{\"u}diger and Valentin, Sascha R and Wieck, Andreas D and Ludwig, Arne and Warburton, Richard J},
  journal={Nature Commun.},
  volume={14},
  number={1},
  pages={3977},
  year={2023},
  url={https://www.nature.com/articles/s41467-023-39568-1}
}

@article{yatsiv1968enhanced,
  title={Enhanced two-proton emission},
  author={Yatsiv, Shaul and Rokni, M and Barak, S},
  journal={Phys. Rev. Lett.},
  volume={20},
  number={23},
  pages={1282},
  year={1968},
  url={https://journals.aps.org/prl/abstract/10.1103/PhysRevLett.20.1282}
}

@article{braunlich1970detection,
  title={Detection of singly stimulated two-photon emission from metastable deuterium atoms},
  author={Br{\"a}unlich, P and Lambropoulos, P},
  journal={Phys. Rev. Lett.},
  volume={25},
  number={3},
  pages={135},
  year={1970},
  url={https://journals.aps.org/prl/abstract/10.1103/PhysRevLett.25.135}
}

@article{malinovsky2001general,
  title={General theory of population transfer by adiabatic rapid passage with intense, chirped laser pulses},
  author={Malinovsky, VS and Krause, JL},
  journal={Eur. Phys. J. D.},
  volume={14},
  number={2},
  pages={147--155},
  year={2001},
  url={https://link.springer.com/article/10.1007/s100530170212}
}

@article{simon2011robust,
  title={Robust Quantum Dot Exciton Generation via Adiabatic Passage with Frequency-Swept Optical Pulses},
  author={Simon, C-M and Belhadj, Thomas and Chatel, B{\'e}atrice and Amand, Thierry and Renucci, Pierre and Lema{\^\i}tre, Aristide and Krebs, Olivier and Dalgarno, PA and Warburton, RJ and Marie, Xavier and others},
  journal={Phys. Rev. Lett.},
  volume={106},
  number={16},
  pages={166801},
  year={2011},
  url={https://journals.aps.org/prl/abstract/10.1103/PhysRevLett.106.166801}
}

@article{wu2011population,
  title={Population Inversion in a Single InGaAs Quantum Dot Using<? format?> the Method of Adiabatic Rapid Passage},
  author={Wu, Yanwen and Piper, IM and Ediger, M and Brereton, P and Schmidgall, ER and Eastham, PR and Hugues, M and Hopkinson, M and Phillips, RT},
  journal={Phys. Rev. Lett.},
  volume={106},
  number={6},
  pages={067401},
  year={2011},
  url={https://journals.aps.org/prl/abstract/10.1103/PhysRevLett.106.067401}
}

@article{kappe2025chirped,
  title={Chirped pulses meet quantum dots: innovations, challenges, and future perspectives},
  author={Kappe, Florian and Karli, Yusuf and Wilbur, Grant and Kr{\"a}mer, Ria G and Ghosh, Sayan and Schwarz, Ren{\'e} and Kaiser, Moritz and Bracht, Thomas K and Reiter, Doris E and Nolte, Stefan and others},
  journal={Adv. Quantum Technol.},
  volume={8},
  number={2},
  pages={2300352},
  year={2025},
  url={https://advanced.onlinelibrary.wiley.com/doi/full/10.1002/qute.202300352}
}

@article{hogg2025fast,
  title={Fast optical control of a coherent hole spin in a microcavity},
  author={Hogg, Mark R and Antoniadis, Nadia O and Marczak, Malwina A and Nguyen, Giang N and Baltisberger, Timon L and Javadi, Alisa and Schott, R{\"u}diger and Valentin, Sascha R and Wieck, Andreas D and Ludwig, Arne and others},
  journal={Nature Physics},
  volume={21},
  number={9},
  pages={1475--1481},
  year={2025},
  url={https://www.nature.com/articles/s41567-025-02988-5}
}

@article{Brion2007,
  title = {Adiabatic elimination in a lambda system},
  volume = {40},
  ISSN = {1751-8121},
  url = {http://dx.doi.org/10.1088/1751-8113/40/5/011},
  DOI = {10.1088/1751-8113/40/5/011},
  number = {5},
  journal = {J. Phys. A},
  publisher = {IOP Publishing},
  author = {Brion,  E and Pedersen,  L H and Mølmer,  K},
  year = {2007},
  month = jan,
  pages = {1033–1043}
}

@article{Bukov2015,
  title = {Universal high-frequency behavior of periodically driven systems: from dynamical stabilization to {Floquet} engineering},
  volume = {64},
  ISSN = {1460-6976},
  url = {http://dx.doi.org/10.1080/00018732.2015.1055918},
  DOI = {10.1080/00018732.2015.1055918},
  number = {2},
  journal = {Adv. Phys.},
  publisher = {Informa UK Limited},
  author = {Bukov,  Marin and D’Alessio,  Luca and Polkovnikov,  Anatoli},
  year = {2015},
  month = mar,
  pages = {139–226}
}

@article{SchimpfPRXQ2025,
  title = {Optical and Magnetic Response by Design in $\mathrm{Ga}\mathrm{As}$ Quantum Dots},
  author = {Schimpf, Christian and Garcia, Ailton J. and Koong, Zhe X. and Nguyen, Giang N. and Niekamp, Lukas L. and Appel, Martin Hayhurst and Hassanen, Ahmed and Waller, James and Karli, Yusuf and Covra da Silva, Saimon Filipe and Ritzmann, Julian and Babin, Hans-Georg and Wieck, Andreas D. and Pishchagin, Anton and Hease, William and Margaria, Nico and Au, Ti-Huong and Boissier, Sebastien and Morassi, Martina and Lemaitre, Aristide and Senellart, Pascale and Somaschi, Niccolo and Ludwig, Arne and Warburton, Richard J. and Atat\"ure, Mete and Rastelli, Armando and Gawe\l{}czyk, Micha\l{} and Gangloff, Dorian A.},
  journal = {PRX Quantum},
  volume = {6},
  issue = {4},
  pages = {040309},
  numpages = {15},
  year = {2025},
  month = {Oct},
  publisher = {American Physical Society},
  doi = {10.1103/98cp-1k42},
  url = {https://link.aps.org/doi/10.1103/98cp-1k42}
}

@article{Shevchenko2010,
  title = {{Landau–Zener–St\"{u}ckelberg interferometry}},
  volume = {492},
  ISSN = {0370-1573},
  DOI = {10.1016/j.physrep.2010.03.002},
  number = {1},
  journal = {Phys. Rep.},
  publisher = {Elsevier BV},
  author = {Shevchenko,  S.N. and Ashhab,  S. and Nori,  Franco},
  year = {2010},
  month = jul,
  pages = {1–30}
}

@article{koong2025coherent,
  title={Coherent Control of Quantum-Dot Spins with Cyclic Optical Transitions},
  author={Koong, Zhe Xian and Haeusler, Urs and Kaspari, Jan M and Schimpf, Christian and Dejen, Benyam and Hassanen, Ahmed M and Graham, Daniel and Garcia Jr, Ailton J and Peter, Melina and Clarke, Edmund and others},
  journal={arXiv preprint arXiv:2509.14445},
  year={2025},
  url = {https://arxiv.org/abs/2509.14445}
}

@article{Golter2023,
author = {Golter, D. Andrew and Clark, Genevieve and El Dandachi, Tareq and Krastanov, Stefan and Leenheer, Andrew J. and Wan, Noel H. and Raniwala, Hamza and Zimmermann, Matthew and Dong, Mark and Chen, Kevin C. and Li, Linsen and Eichenfield, Matt and Gilbert, Gerald and Englund, Dirk},
title = {Selective and Scalable Control of Spin Quantum Memories in a Photonic Circuit},
journal = {Nano Lett.},
volume = {23},
number = {17},
pages = {7852-7858},
year = {2023},
URL = {https://doi.org/10.1021/acs.nanolett.3c01511}
}

@article{
Gangloff2019,
author = {D. A. Gangloff  and G. Éthier-Majcher  and C. Lang  and E. V. Denning  and J. H. Bodey  and D. M. Jackson  and E. Clarke  and M. Hugues  and C. Le Gall  and M. Atatüre },
title = {Quantum interface of an electron and a nuclear ensemble},
journal = {Science},
volume = {364},
number = {6435},
pages = {62-66},
year = {2019},
doi = {10.1126/science.aaw2906},
URL = {https://www.science.org/doi/abs/10.1126/science.aaw2906},
}

@article{Chekhovich2020,
	author = {Chekhovich, Evgeny A. and da Silva, Saimon F. Covre and Rastelli, Armando},
	date = {2020/12/01},
	doi = {10.1038/s41565-020-0769-3},
	id = {Chekhovich2020},
	isbn = {1748-3395},
	journal = {Nat. Nanotechnol.},
	number = {12},
	pages = {999--1004},
	title = {Nuclear spin quantum register in an optically active semiconductor quantum dot},
	url = {https://doi.org/10.1038/s41565-020-0769-3},
	volume = {15},
	year = {2020}}

@article{Shofer2025,
  title = {Tuning the Coherent Interaction of an Electron Qubit and a Nuclear Magnon},
  author = {Shofer, Noah and Zaporski, Leon and Hayhurst Appel, Martin and Manna, Santanu and Covre da Silva, Saimon and Ghorbal, Alexander and Haeusler, Urs and Rastelli, Armando and Le Gall, Claire and Gawe\l{}czyk, Micha\l{} and Atat\"ure, Mete and Gangloff, Dorian A.},
  journal = {Phys. Rev. X},
  volume = {15},
  issue = {2},
  pages = {021004},
  numpages = {21},
  year = {2025},
  month = {Apr},
  publisher = {American Physical Society},
  doi = {10.1103/PhysRevX.15.021004},
  url = {https://link.aps.org/doi/10.1103/PhysRevX.15.021004}
}

@article{Appel2025,
	author = {Appel, Martin Hayhurst and Ghorbal, Alexander and Shofer, Noah and Zaporski, Leon and Manna, Santanu and da Silva, Saimon Filipe Covre and Haeusler, Urs and Le Gall, Claire and Rastelli, Armando and Gangloff, Dorian A. and Atat{\"u}re, Mete},
	date = {2025/03/01},
	doi = {10.1038/s41567-024-02746-z},
	id = {Appel2025},
	isbn = {1745-2481},
	journal = {Nature Physics},
	number = {3},
	pages = {368--373},
	title = {A many-body quantum register for a spin qubit},
	url = {https://doi.org/10.1038/s41567-024-02746-z},
	volume = {21},
	year = {2025}}

@article{Zaporski2023,
	author = {Zaporski, Leon and Shofer, Noah and Bodey, Jonathan H. and Manna, Santanu and Gillard, George and Appel, Martin Hayhurst and Schimpf, Christian and Covre da Silva, Saimon Filipe and Jarman, John and Delamare, Geoffroy and Park, Gunhee and Haeusler, Urs and Chekhovich, Evgeny A. and Rastelli, Armando and Gangloff, Dorian A. and Atat{\"u}re, Mete and Le Gall, Claire},
	date = {2023/03/01},
	doi = {10.1038/s41565-022-01282-2},
	id = {Zaporski2023},
	isbn = {1748-3395},
	journal = {Nat. Nanotechnol.},
	number = {3},
	pages = {257--263},
	title = {Ideal refocusing of an optically active spin qubit under strong hyperfine interactions},
	url = {https://doi.org/10.1038/s41565-022-01282-2},
	volume = {18},
	year = {2023}}

@article{dyte2025,
      title={Storing quantum coherence in a quantum dot nuclear spin ensemble for over 100 milliseconds}, 
      author={Harry E. Dyte and Santanu Manna and Saimon F. Covre da Silva and Armando Rastelli and Evgeny A. Chekhovich},
      year={2026},
      journal = {Nat. Commun.},
      volume = {17},
      pages = {239},
      url={https://www.nature.com/articles/s41467-025-66948-6}, 
}

@article{Gazzano2013,
	author = {Gazzano, O. and Michaelis de Vasconcellos, S. and Arnold, C. and Nowak, A. and Galopin, E. and Sagnes, I. and Lanco, L. and Lema{\^\i}tre, A. and Senellart, P.},
	date = {2013/02/05},
	doi = {10.1038/ncomms2434},
	id = {Gazzano2013},
	isbn = {2041-1723},
	journal = {Nat. Commun.},
	number = {1},
	pages = {1425},
	title = {Bright solid-state sources of indistinguishable single photons},
	url = {https://doi.org/10.1038/ncomms2434},
	volume = {4},
	year = {2013}}

@article{Holewa2022,
	author = {Holewa, Pawe{\l} and Sakanas, Aurimas and G{\"u}r, Ugur M. and Mrowi{\'n}ski, Pawe{\l} and Huck, Alexander and Wang, Bi-Ying and Musia{\l}, Anna and Yvind, Kresten and Gregersen, Niels and Syperek, Marcin and Semenova, Elizaveta},
	date = {2022/07/20},
	doi = {10.1021/acsphotonics.2c00027},
	journal = {ACS Photonics},
	month = {07},
	number = {7},
	pages = {2273--2279},
	publisher = {American Chemical Society},
	title = {Bright Quantum Dot Single-Photon Emitters at Telecom Bands Heterogeneously Integrated on Si},
	type = {doi: 10.1021/acsphotonics.2c00027},
	url = {https://doi.org/10.1021/acsphotonics.2c00027},
	volume = {9},
	year = {2022},
	year1 = {2022}}

@article{Senellart2017,
	author = {Senellart, Pascale and Solomon, Glenn and White, Andrew},
	date = {2017/11/01},
	doi = {10.1038/nnano.2017.218},
	id = {Senellart2017},
	isbn = {1748-3395},
	journal = {Nat. Nanotechnol.},
	number = {11},
	pages = {1026--1039},
	title = {High-performance semiconductor quantum-dot single-photon sources},
	url = {https://doi.org/10.1038/nnano.2017.218},
	volume = {12},
	year = {2017}}

@article{Pingault2017,
	author = {Pingault, Benjamin and Jarausch, David-Dominik and Hepp, Christian and Klintberg, Lina and Becker, Jonas N. and Markham, Matthew and Becher, Christoph and Atat{\"u}re, Mete},
	date = {2017/05/30},
	doi = {10.1038/ncomms15579},
	id = {Pingault2017},
	isbn = {2041-1723},
	journal = {Nat. Commun.},
	number = {1},
	pages = {15579},
	title = {Coherent control of the silicon-vacancy spin in diamond},
	url = {https://doi.org/10.1038/ncomms15579},
	volume = {8},
	year = {2017}}

@article{Debroux2021,
  title = {Quantum Control of the Tin-Vacancy Spin Qubit in Diamond},
  author = {Debroux, Romain and Michaels, Cathryn P. and Purser, Carola M. and Wan, Noel and Trusheim, Matthew E. and Arjona Mart\'{\i}nez, Jes\'us and Parker, Ryan A. and Stramma, Alexander M. and Chen, Kevin C. and de Santis, Lorenzo and Alexeev, Evgeny M. and Ferrari, Andrea C. and Englund, Dirk and Gangloff, Dorian A. and Atat\"ure, Mete},
  journal = {Phys. Rev. X},
  volume = {11},
  issue = {4},
  pages = {041041},
  numpages = {9},
  year = {2021},
  month = {Nov},
  publisher = {American Physical Society},
  doi = {10.1103/PhysRevX.11.041041},
  url = {https://link.aps.org/doi/10.1103/PhysRevX.11.041041}
}

@article{Hepp2014,
  title = {Electronic Structure of the Silicon Vacancy Color Center in Diamond},
  author = {Hepp, Christian and M\"uller, Tina and Waselowski, Victor and Becker, Jonas N. and Pingault, Benjamin and Sternschulte, Hadwig and Steinm\"uller-Nethl, Doris and Gali, Adam and Maze, Jeronimo R. and Atat\"ure, Mete and Becher, Christoph},
  journal = {Phys. Rev. Lett.},
  volume = {112},
  issue = {3},
  pages = {036405},
  numpages = {5},
  year = {2014},
  month = {Jan},
  publisher = {American Physical Society},
  doi = {10.1103/PhysRevLett.112.036405},
  url = {https://link.aps.org/doi/10.1103/PhysRevLett.112.036405}
}

@article{samaras2026,
      title={Complete coherent control of spin qubits in self-assembled InAs quantum dots under oblique magnetic fields}, 
      author={I. Samaras and K. Barr and C. Schneider and S. Höfling and K. G. Lagoudakis},
      year={2026},
      journal={2604.07074},
      url={https://arxiv.org/abs/2604.07074}
}

@article{Dutt2007a,
   author = {M. V. G. Dutt and L. Childress and L. Jiang and E. Togan and J. Maze and F. Jelezko and A. S. Zibrov and P. R. Hemmer and M. D. Lukin},
   doi = {10.1126/science.1139831},
   issn = {0036-8075},
   issue = {5829},
   journal = {Science},
   month = {6},
   pages = {1312-1316},
   title = {Quantum Register Based on Individual Electronic and Nuclear Spin Qubits in Diamond},
   volume = {316},
   url = {http://www.sciencemag.org/cgi/doi/10.1126/science.1139831},
   year = {2007}
}

@article{Simon2007a,
   author = {Jonathan Simon and Haruka Tanji and Saikat Ghosh and Vladan Vuletić},
   doi = {10.1038/nphys726},
   issn = {1745-2473},
   issue = {11},
   journal = {Nat. Phys.},
   month = {9},
   pages = {765-769},
   title = {Single-photon bus connecting spin-wave quantum memories},
   volume = {3},
   url = {http://www.nature.com/doifinder/10.1038/nphys726},
   year = {2007}
}

@article{Moehring2007a,
   author = {D. L. Moehring and P. Maunz and S. Olmschenk and K. C. Younge and D. N. Matsukevich and L.-M. Duan and C. Monroe},
   doi = {10.1038/nature06118},
   issn = {0028-0836},
   issue = {7158},
   journal = {Nature},
   month = {9},
   pages = {68-71},
   title = {Entanglement of single-atom quantum bits at a distance},
   volume = {449},
   url = {http://www.nature.com/articles/nature06118},
   year = {2007}
}

@article{Gao2012,
   author = {W. B. Gao and P. Fallahi and E. Togan and J. Miguel-Sanchez and A. Imamoglu},
   doi = {10.1038/nature11573},
   issn = {0028-0836},
   issue = {7424},
   journal = {Nature},
   month = {11},
   pages = {426-430},
   publisher = {Nature Research},
   title = {Observation of entanglement between a quantum dot spin and a single photon},
   volume = {491},
   url = {http://www.nature.com/doifinder/10.1038/nature11573 http://www.nature.com/articles/nature11573},
   year = {2012}
}

@article{Greilich2006,
   author = {A. Greilich and D. R. Yakovlev and A. Shabaev and Al. L. Efros and I. A. Yugova and R. Oulton and V. Stavarache and D. Reuter and A. Wieck and M. Bayer},
   doi = {10.1126/science.1128215},
   issn = {0036-8075},
   issue = {5785},
   journal = {Science},
   month = {7},
   pages = {341-345},
   title = {Mode Locking of Electron Spin Coherences in Singly Charged Quantum Dots},
   volume = {313},
   url = {https://www.science.org/doi/10.1126/science.1128215},
   year = {2006}
}

@article{xu2009,
   author = {Xiaodong Xu and Wang Yao and Bo Sun and Duncan G Steel and Allan S Bracker and Daniel Gammon and L J Sham},
   doi = {10.1038/nature08120},
   isbn = {978-1-55752-869-8},
   issn = {0028-0836},
   issue = {7250},
   journal = {Nature},
   month = {6},
   note = {10.1038/nature08120},
   pages = {1105-1109},
   pmid = {19553994},
   publisher = {Macmillan Publishers Limited. All rights reserved},
   title = {Optically controlled locking of the nuclear field via coherent dark-state spectroscopy},
   volume = {459},
   url = {http://dx.doi.org/10.1038/nature08120 http://www.nature.com/doifinder/10.1038/nature08120},
   year = {2009}
}

@article{Hogele2012a,
   author = {A. Högele and M. Kroner and C. Latta and M. Claassen and I. Carusotto and C. Bulutay and A. Imamoglu},
   doi = {10.1103/PhysRevLett.108.197403},
   issn = {0031-9007},
   issue = {19},
   journal = {Phys. Rev. Lett.},
   month = {5},
   pages = {197403},
   title = {Dynamic Nuclear Spin Polarization in the Resonant Laser Excitation of an InGaAs Quantum Dot},
   volume = {108},
   url = {https://link.aps.org/doi/10.1103/PhysRevLett.108.197403},
   year = {2012}
}

@article{Bodey2019,
   author = {J. H. Bodey and R. Stockill and E. V. Denning and D. A. Gangloff and G. Éthier-Majcher and D. M. Jackson and E. Clarke and M. Hugues and C. Le Gall and M. Atatüre},
   doi = {10.1038/s41534-019-0206-3},
   issn = {2056-6387},
   issue = {1},
   journal = {npj Quantum Information},
   month = {12},
   pages = {95},
   title = {Optical spin locking of a solid-state qubit},
   volume = {5},
   url = {http://arxiv.org/abs/1906.00427 http://www.nature.com/articles/s41534-019-0206-3},
   year = {2019}
}

@article{Jackson2021,
   author = {Daniel M. Jackson and Dorian A. Gangloff and Jonathan H. Bodey and Leon Zaporski and Clara Bachorz and Edmund Clarke and Maxime Hugues and C. Le Gall and Mete Atatüre},
   doi = {10.1038/s41567-020-01161-4},
   issn = {1745-2473},
   issue = {5},
   journal = {Nat. Phys.},
   month = {5},
   pages = {585-590},
   title = {Quantum sensing of a coherent single spin excitation in a nuclear ensemble},
   volume = {17},
   url = {http://arxiv.org/abs/2008.09541 http://www.nature.com/articles/s41567-020-01161-4},
   year = {2021}
}

\end{document}